\documentclass[english,aps,preprint,superscriptaddress]{revtex4-1}
\usepackage{mathptmx}
\usepackage[T1]{fontenc}
\usepackage[latin9]{inputenc}
\usepackage{amsmath}
\usepackage{amssymb}
\usepackage{graphicx}
\usepackage{esint}

\makeatletter
\renewcommand{\fnum@figure}{\textbf{Figure~\thefigure}}

\setcitestyle{super}

\makeatother

\usepackage{babel}
\begin{document}

\title{Dynamic detection of electron spin accumulation in ferromagnet-semiconductor
devices by ferromagnetic resonance}

\author{Changjiang~Liu}

\affiliation{School of Physics and Astronomy, University of Minnesota, Minneapolis,
Minnesota 55455,~USA}

\author{Sahil~J.~Patel}

\affiliation{Department of Materials, University of California, Santa Barbara,
California 93106,~USA}

\author{Timothy~A.~Peterson}

\affiliation{School of Physics and Astronomy, University of Minnesota, Minneapolis,
Minnesota 55455,~USA}

\author{Chad~C.~Geppert}

\affiliation{School of Physics and Astronomy, University of Minnesota, Minneapolis,
Minnesota 55455,~USA}

\author{Kevin~D.~Christie}

\affiliation{School of Physics and Astronomy, University of Minnesota, Minneapolis,
Minnesota 55455,~USA}

\author{Gordon~Stecklein}

\affiliation{School of Physics and Astronomy, University of Minnesota, Minneapolis,
Minnesota 55455,~USA}

\author{Chris~J.~Palmstr\o m}

\affiliation{Department of Materials, University of California, Santa Barbara,
California 93106,~USA}

\affiliation{Department of Electrical and Computer Engineering, University of
California, Santa Barbara, California 93106,~USA}

\author{Paul~A.~Crowell}

\email{crowell@umn.edu}

\selectlanguage{english}%

\affiliation{School of Physics and Astronomy, University of Minnesota, Minneapolis,
Minnesota 55455,~USA}
\begin{abstract}
A distinguishing feature of spin accumulation in ferromagnet-semiconductor
devices is precession of the non-equilibrium spin population of the
semiconductor in a magnetic field. This is the basis for detection
techniques such as the Hanle effect, but these approaches become less
effective as the spin lifetime in the semiconductor decreases. For
this reason, no electrical Hanle measurement has been demonstrated
in GaAs at room temperature. We show here that by forcing the magnetization
in the ferromagnet (the spin injector and detector) to precess at
the ferromagnetic resonance frequency, an electrically generated spin
accumulation can be detected from 30 to 300~K. At low temperatures,
the distinct Larmor precession of the spin accumulation in the semiconductor
can be detected by ferromagnetic resonance in an oblique field. We
verify the effectiveness of this new spin detection technique by comparing
the injection bias and temperature dependence of the measured spin
signal to the results obtained using traditional methods. We further
show that this new approach enables a measurement of short spin lifetimes
(< 100~psec), a regime that is not accessible in semiconductors using
traditional Hanle techniques.
\end{abstract}
\maketitle
An effective means to electrically detect electron spin accumulation
in a semiconductor is essential to the development of semiconductor
spintronics\cite{Chappert2007,Dery2006}. A common detection method,
known as the Hanle effect\cite{F.Meier1984}, relies on the precession
of the spin accumulation in the semiconductor by an external magnetic
field. This approach has been employed in both the non-local \cite{Jedema2002,Lou2007}
and the local 3-terminal (3T) \cite{Lou2006} measurement geometries.
As the spin lifetime decreases at higher temperatures, a significantly
higher magnetic field is required in order for the Hanle effect to
dephase the spin accumulation\cite{Ziutifmmodecuteclseci2004}. The
ordinary magnetoresistance in semiconductors in such large fields
produces large quadratic backgrounds\cite{Lou2006} that mask any
truly spin-dependent effects. As a result, no electrical Hanle measurement
has been demonstrated in $n$-GaAs at room temperature. Although the
3T technique has recently been widely used to study spin transport
in new material systems\cite{Dash2009,Li2011,Swartz2014} at higher
temperatures, it has been shown that the 3T Hanle measurement is sensitive
to a variety of magnetic field assisted phenomena\cite{Tinkey2014,Txoperena2014,Song2014},
which often cannot be separated from the Hanle effect.

We introduce here a new detection technique, which utilizes the precession
of the magnetization under ferromagnetic resonance (FMR) to dynamically
detect the spin accumulation in a semiconductor. The approach is similar
in some aspects to a Hanle measurement, in that the observed signal
corresponds to a suppression of the component of the spin accumulation
parallel to the magnetization. In this case, however, the suppression
is due to the onset of FMR rather than precession of the spin accumulation
in an orthogonal magnetic field. The narrow FMR linewidth and the
sensitivity to both the precession of the magnetization in the ferromagnet
as well as the dynamics of the spin accumulation in the semiconductor
provide an immunity to the field-dependent backgrounds that affect
Hanle measurements, particularly in the 3T case. At low temperatures,
the Larmor precession of the spin accumulation in the semiconductor
can be detected by an FMR measurement in an oblique magnetic field.
We verify the effectiveness of this approach by comparing the bias
and temperature dependence of the measured spin signal to results
obtained on the same heterostructures using traditional spin detection
methods. Finally, we show that this FMR-based spin detection technique
enables one to determine the spin relaxation rate in GaAs at room
temperature, at which an ordinary Hanle curve cannot be measured. 

The devices used in this electrical spin injection and detection experiment
consist of ferromagnet (FM)$/$semiconductor (SC) heterostructures.
A spin current across the FM$/$SC interface is generated by a current
source\cite{Hanbicki2002,Motsnyi2002}, leading to a spin accumulation,
which has previously been detected potentiometrically\cite{Johnson1985,Lou2007}
or by using a spin filter to detect the polarized current\cite{Appelbaum2007}.
In the potentiometric approach, the voltage detected by a FM contact
due to spin accumulation can be expressed as\cite{Lou2007}
\begin{equation}
V=\eta P_{\mathrm{FM}}\dfrac{1}{e}\frac{\partial\mu}{\partial n}\mathbf{s\cdot\hat{m}},\label{eq:1}
\end{equation}
where $\eta$ is the spin detection efficiency, $P_{\mathrm{FM}}$
is the spin polarization at the Fermi level in the FM, $e$ is the
electron charge, and $\mu$ and $n$ are the chemical potential and
number density of electrons in the SC. The spin accumulation $\mathbf{s}$
in the SC has a magnitude defined as $\mathbf{\left|s\right|=\mathrm{\mathit{n}}_{\uparrow}-\mathrm{\mathit{n}}_{\downarrow}},$
where $\mathrm{\mathit{n}}_{\uparrow}$ and $\mathrm{\mathit{n}}_{\downarrow}$
refer to the number density of spin-up and spin-down electrons, and
$\mathbf{\hat{m}}$ is the unit vector of the magnetization $\mathbf{m}$
in the FM. The dot product in equation~(\ref{eq:1}) accounts for
the projection of the spin accumulation, which must be averaged over
the area of the contact, onto the magnetization. In a Hanle measurement,
one applies a magnetic field perpendicular to $\mathbf{\hat{m}}$
to precess and dephase the injected spins, thus reducing the magnitude
and changing the orientation of $\mathbf{s}$. In our experiment,
we force the magnetization of the FM to precess instead of the spin
accumulation in the SC. We inject spins from the FM and simultaneously
drive the magnetization in the FM to precess by FMR. During this process,
$\mathbf{\mathbf{s\cdot\hat{m}}}$ also changes, which allows for
detection of the spin accumulation. As will be explained in detail
in the modeling section, the important time scales in this experiment
are the FMR precession period and the electron spin lifetime $\tau_{s}$
in the SC. Together these determine the steady state value of $\mathbf{\mathbf{s\cdot\hat{m}}}$,
which changes at resonance.

\noindent 
\begin{figure}[!h]
\includegraphics{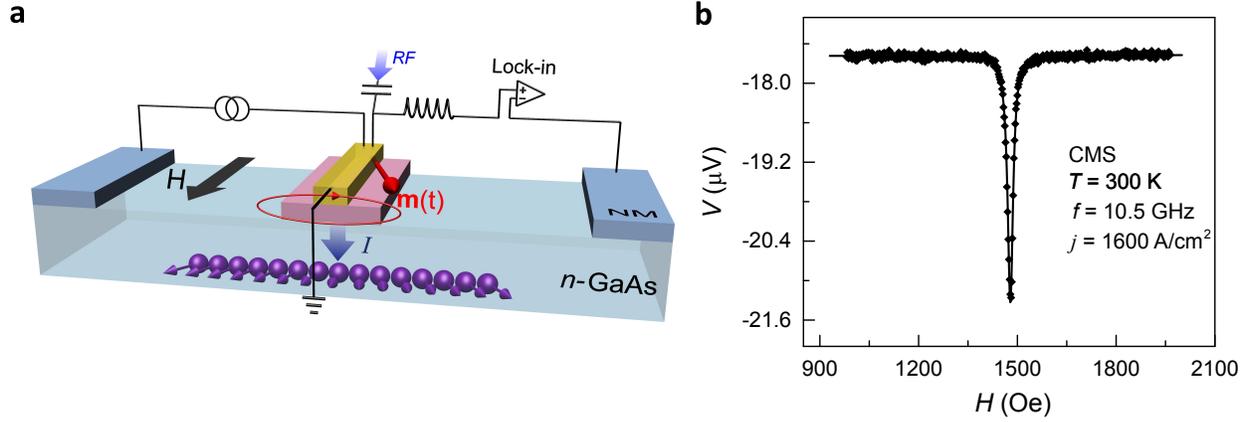}

\protect\caption{\textbf{Schematic diagram of the experiment and a representative plot
of the spin accumulation data measured by FMR at room temperature.
}(\textbf{a}) A schematic diagram of the FMR-spin detection experiment.
The red contact in the middle of the device is ferromagnetic. On top
of the ferromagnet is the microwave stripline (shown in yellow). The
two remote contacts (shown in blue) are ohmic and are fabricated from
non-magnetic Cu-Ge\cite{Aboelfotoh1994}. $\mathit{\mbox{\ensuremath{\mathbf{H}}}}$
and $\mathit{\mathbf{m}\mathrm{(t)}}$ represent the applied magnetic
field and the precessing magnetization of the FM, respectively. \textit{I}
is the applied interface bias current, which generates a spin accumulation
in $n$-GaAs. Spins (shown in purple) injected by the precessing magnetization
at different times have different orientations, and they diffuse in
the $n$-GaAs channel. (\textbf{b}) Spin signal measured by FMR as
a decrease in voltage at room temperature. The solid line is a Lorentzian
fit to the data. CMS refers to the sample with \textit{$\mathrm{Co{}_{2}MnSi}$
}as the FM\textit{.\label{fig:1}}}
\end{figure}
\textbf{\large{}Results}{\large \par}

\noindent \textbf{Experimental setup and room temperature measurement.}
Figure~\ref{fig:1}a is a schematic diagram of the FMR-spin detection
experiment. The devices used in our experiment are fabricated from
epitaxial FM\textit{$/n$-}GaAs (100) heterostructures. Recently,
the utilization of the ferromagnetic Heusler alloys\cite{Wang2005}
has greatly improved the spin injection efficiency into GaAs\cite{Akiho2013,Saito2013}.
In our experiment, we use the Heusler alloys \textit{$\mathrm{Co{}_{2}MnSi}$}
and $\mathrm{Co{}_{2}FeSi}$ as spin injectors. The FM film is 5 nm
thick with lateral dimensions of $5\times50$~$\mathrm{\mu m}$.
The $n$-GaAs is doped with a concentration of $n=3\times10^{16}$
$\mathrm{cm^{-3}}$. The junction region consists of a highly doped
$n^{+}(5\times10^{18}$ $\mathrm{cm^{-3}})$ layer to thin the Schottky
tunnel barrier in order to achieve higher spin injection efficiency\cite{Rashba2000,Schmidt2000}.
A 120~nm thick gold wire, which forms a microwave stripline, is deposited
directly on top of the FM contact. The microwave Oersted field from
the stripline drives the resonance. The microwave frequency used in
the experiment ranges from 5~Ghz to 20~Ghz. The applied dc magnetic
field is in the sample plane as indicated by $\mathbf{H}$ in Fig.~\ref{fig:1}a.
A forward bias current across the FM$/$SC interface injects a spin
current into the \textit{$n$-}GaAs\cite{Stephens2004}. The microwave
field is modulated at low frequency (100~Hz) and a lock-in amplifier
synchronized with the modulation frequency measures the difference
in interface voltage with microwaves on and off.

When the bias current generates a spin accumulation in the \textit{$n$-}GaAs
and the magnetization is driven on resonance, a dip in the three-terminal
voltage is observed. The raw data in Figure~\ref{fig:1}b show the
resonance observed at room temperature. The resonance field measured
as a function of the microwave frequency agrees with the FMR spectrum
as calculated from Kittel's formula\cite{Kittel1948} using the anisotropy
and magnetization of each sample (see Supplementary Fig.~1). The
non-zero background voltage in the plot is due to the nonlinear IV
characteristic of the FM$/n$-GaAs Schottky contact\cite{Sharma1984},
which rectifies the microwave current generated in the device. We
see that the interface voltage decreases on resonance, which, as will
be explained in detail below, is the result of the decrease of the
projection of $\mathbf{s}$ onto $\mathbf{\hat{m}}$. In previous
work, we found that the tunneling anisotropic magnetoresistance (TAMR)
effect at the FM$/$SC interface also produces a change in the interface
voltage at FMR\cite{Liu2014}. The size of the TAMR effect depends
on the bias voltage across the FM$/$SC interface\cite{Moser2007,Liu2014}.
When a high forward bias current is applied to the sample generating
a spin accumulation in the SC, the FMR signal due to the TAMR effect
is much smaller than that produced by the spin accumulation\cite{Liu2014}.
In the data analysis below, the small contribution to the FMR signal
from the TAMR effect has been removed.

\noindent 
\begin{figure}[!h]
\includegraphics{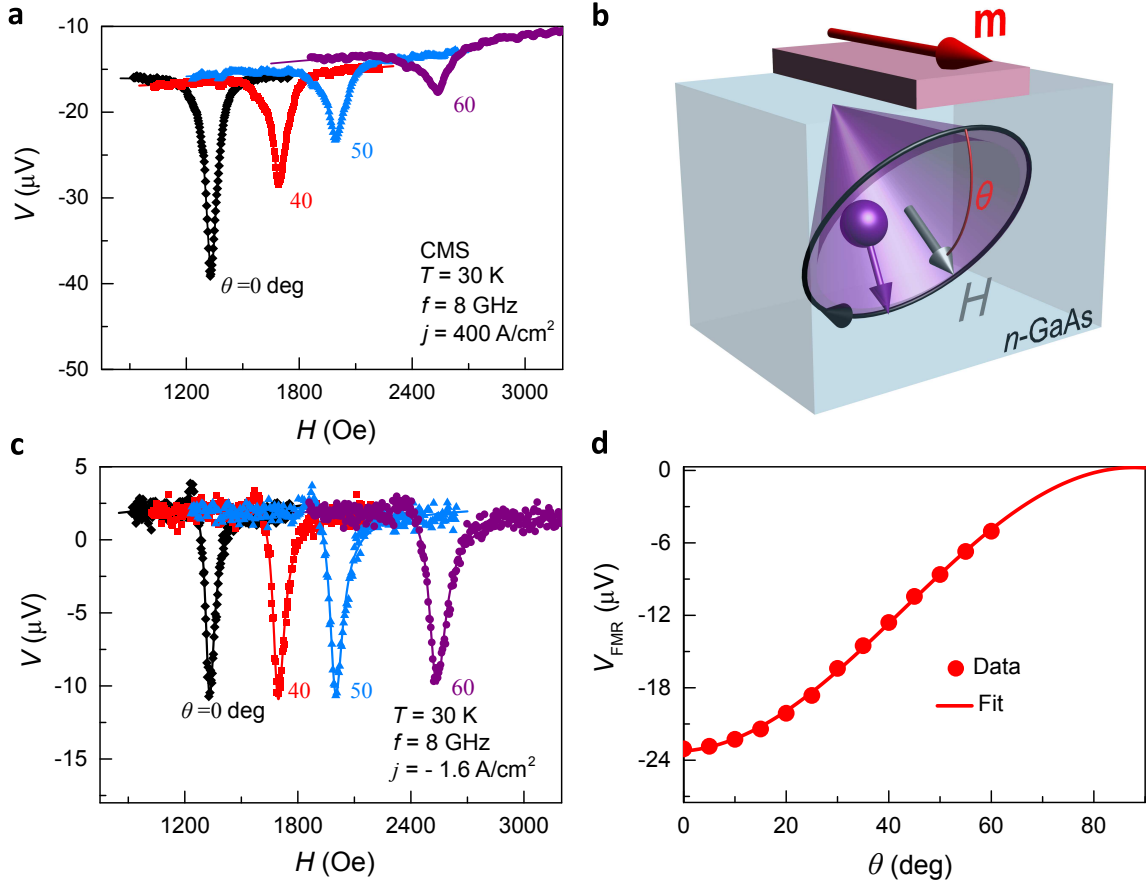}

\protect\caption{\textbf{Experimental results in the oblique configuration at 30 K
and illustration of the Larmor precession of the spin accumulation.}
(\textbf{a}) The FMR signal measured at different out-of-plane angles
of the applied field when a spin accumulation exists in the \textit{$n$-}GaAs.
Note the decrease in the signal amplitude with increasing angle. (\textbf{b})
The applied magnetic field $\mathit{\mathbf{H}}$, represented by
the grey arrow, is tilted out-of-plane by an angle $\theta$. The
injected spins precess about $\mathit{\mathbf{H}}$ and their directions
are uniformly distributed on a cone when $\tau_{s}\gg\frac{2\pi}{\omega_{L}}$,
as is the case at 30 K. (\textbf{c}) FMR measured under reverse bias,
at which only the TAMR effect is present. The out-of-plane angles
are the same as in Fig.~\ref{fig:2}a. In this case the amplitude
is independent of angle. (\textbf{d}) Angle dependence of the magnitude
of the spin accumulation signal measured by FMR (Fig.~\ref{fig:2}a),
and a fit based on the Larmor precession model. Because of the large
demagnetizing field, the precession axis of the magnetization remains
nearly in the sample plane. The raw data in (\textbf{a},\textbf{c})
are plotted as a function of the total magnetic field, and so the
resonance field positions and the widths of the peaks scale as $1/\cos\theta$
relative to the in-plane ($\theta=0$) case. \label{fig:2}}
\end{figure}
\textbf{Detecting the Larmor precession.} A distinguishing feature
of spin is the Larmor precession about an applied magnetic field.
The precession frequency is $\omega_{L}=g\mu_{B}\mathrm{\mathit{B}}/\hbar$,
where $\mu_{B}$ is the Bohr magneton, the $g$ factor is $-0.44$
for electrons in GaAs\cite{Kato2004} and $\hbar$ is Planck's constant.
We expect that the Larmor precession in the SC will influence the
spin accumulation measured by FMR when $\omega_{L}$ is larger than
the spin relaxation rate. In these samples, this condition can be
achieved at low temperatures (30~K), at which the spin lifetime for
this doping is longer than 10~ns\cite{Furis2006}, while the the
Larmor precession period is about 1 ns for a magnetic field of 2000~Oe.
We tilt the magnetic field out of the sample plane such that the perpendicular
component of the magnetic field is still much smaller than the magnetic
anisotropy field of the FM film. This allows the magnetization to
remain nearly in the sample plane, while the spins in the semiconductor
precess about the oblique magnetic field. In steady state, the spin
accumulation can be viewed as an ensemble of spins injected at different
times\cite{F.Meier1984}. When $\omega_{L}>\tau_{s}^{-1},$ as is
the case at 30~K, the orientations of electron spins are therefore
distributed uniformly on the precession cone, which is depicted in
purple in Fig.~\ref{fig:2}b. In other words, the \emph{transverse
}component of the spin accumulation is completely dephased, and there
is no component of the steady-state spin accumulation perpendicular
to the applied field.

Figure~\ref{fig:2}a shows that the FMR spin accumulation signal
decreases significantly as the out-of-plane angle $\theta$ of the
magnetic field increases. To verify that this change is not simply
due to the magnetization dynamics of the FM, we confirm that the amplitude
of the precession cone angle in the FM does not change significantly
with increasing angle by applying a \emph{reverse} bias across the
FM$/n$-GaAs interface. Under reverse bias, there is no spin accumulation
generated in the \textit{$n$-}GaAs, as we have verified in traditional
non-local spin valve measurements, and the interface voltage at resonance
is due only to the TAMR. The TAMR can in turn be related very simply
to the geometry of the precessing magnetization in the FM (see Supplementary
Note 2). Figure~\ref{fig:2}c shows almost no change in the size
of the voltage peak when the field is tilted out of plane, and hence
we conclude that the amplitude of the precession cone angle in the
FM is nearly constant. 

The spin accumulation detected by the FM is proportional to $\mathbf{\left\langle \mathbf{\mathbf{s\cdot\hat{m}}}\right\rangle }$,
where the brackets indicate the time average in steady state. Because
the transverse spin is completely dephased under the conditions being
considered, the \emph{magnitude }of the spin accumulation is reduced
by a factor of $\cos\theta$ relative to its magnitude $s_{0}$ when
the field is in the sample plane. As a result, $\mathbf{s}=s_{0}\cos\theta\hat{\mathbf{h}}$,
where $\hat{\mathbf{h}}$ is a unit vector parallel to the magnetic
field. Because the detected voltage is proportional to $\mathbf{\left\langle \mathbf{\mathbf{s\cdot\hat{m}}}\right\rangle }$
and none of the prefactors in equation~(\ref{eq:1}) depend on angle,
the spin accumulation signal at resonance should therefore vary as
$\cos^{2}\theta$. Figure~\ref{fig:2}d shows the amplitude of the
resonant signal $V_{\mathrm{FMR}}$ (the magnitudes of the negative
peaks in Fig.~\ref{fig:2}a) as a function of $\theta.$ The solid
line is a fit to $\cos^{2}\theta$, and so the behavior expected for
a spin accumulation signal is observed.

\begin{figure}[!h]
\includegraphics{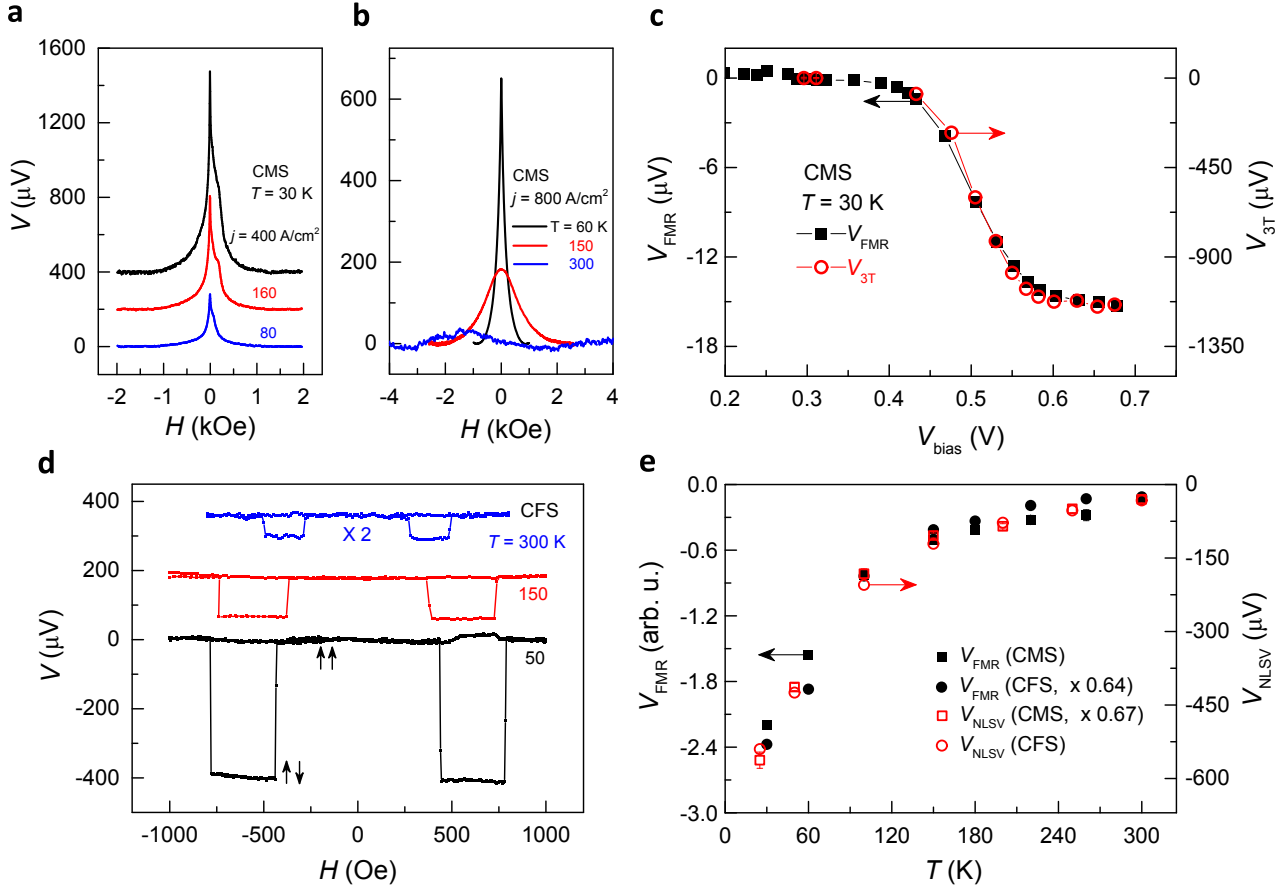}

\protect\caption{\textbf{Injection bias and temperature dependence of the spin accumulation
measured by traditional methods and their comparison to the results
obtained by the FMR technique.} (\textbf{a}) 3T data measured as a
function of perpendicular field at 30 K for different injection bias
currents. (\textbf{b}) 3T signal measured at different temperatures
for a fixed bias current. The spin signal cannot be resolved using
the Hanle measurement at 300 K. (\textbf{c}) Magnitude of the FMR
peak, $V_{\mathrm{FMR}}$ (black squares, left axis), and $V_{\mathrm{3T}}$
(red circles, right axis) plotted as a function of injection bias
voltage at 30 K. Note that the two sets of measurements differ by
only a single bias-independent scale factor. (\textbf{d}) Non-local
spin valve measurements at different temperatures, obtained at a source-detector
separation of 250 nm. The 300 K data are multiplied by a factor of
2 for clarity. The arrows indicate the relative orientations of the
magnetization in the source and detector FM. In \textbf{a},\textbf{
b }and \textbf{d} a second order background has been subtracted from
the raw data. Curves in \textbf{a} and \textbf{d} are offset for clarity.
(\textbf{e}) $V_{\mathrm{FMR}}$ (black solid symbols, left axis)
and $V_{\mathrm{NLSV}}$ (red open symbols, right axis) measured as
a function of temperature. CMS and CFS refer to samples with $\mathrm{Co{}_{2}MnSi}$
and $\mathrm{Co{}_{2}FeSi}$ as the FM, respectively. \label{fig:3}}
\end{figure}

\noindent \textbf{Comparison to other measurement techniques.} To
obtain further evidence that the observed FMR signal is due to the
spin accumulation in the $n$-GaAs layer, we measure the dependence
of $V_{\mathrm{FMR}}$ on the bias current and temperature. The resulting
data can then be compared to those obtained using other techniques.
We first use the 3T Hanle measurement, carried out in a perpendicular
field, to measure the spin accumulation at different injection bias
currents\cite{Lou2006,Crooker2009}. Figure~\ref{fig:3}a shows the
measured 3T signal $V_{\mathrm{3T}}$ at 30~K for different forward
bias currents, and Fig.~\ref{fig:3}b shows $V_{\mathrm{3T}}$ at
a fixed forward bias current for different temperatures. The asymmetric
lineshape and the narrowing of the peak at low magnetic field in Fig.~\ref{fig:3}a
are caused by the hyperfine interaction between the injected spins
and the polarized nuclei\cite{Chan2009}. At low temperature, $V_{\mathrm{3T}}$
is nearly proportional to the spin accumulation\cite{Lou2006}, and
so we expect the bias dependences of $V_{\mathrm{3T}}$ and $V_{\mathrm{FMR}}$
to be the same. In Fig.~\ref{fig:3}c, $V_{\mathrm{3T}}$ and the
FMR signal at 30 K are plotted as a function of the bias voltage.
The two measurement techniques agree up to a single bias-independent
scale factor. Unfortunately, this comparison cannot be made at high
temperatures, at which a spin-dependent contribution to $V_{\mathrm{3T}}$
cannot be measured, as can be seen from the 300 K data in Fig.~\ref{fig:3}b.
We have also performed non-local Hanle measurements\cite{Lou2007}
on this sample (see Supplementary Fig.~2) . Although consistent with
$V_{\mathrm{3T}}$ up to 120~K, the non-local Hanle signal also cannot
be measured at higher temperatures.

The steady-state spin accumulation in the semiconductor depends on
the spin lifetime and diffusion constant \cite{Ziutifmmodecuteclseci2004},
which are strong functions of temperature. The Hanle techniques fail
at high temperatures because the magnetic field scale corresponding
to $\omega_{L}\tau_{s}\sim1$ grows dramatically as $\tau_{s}$ decreases.
As a result, the Hanle curves broaden and cannot be distinguished
from field-dependent backgrounds. This limitation does not impact
the FMR signal, as we can see by comparison with non-local spin valve
measurements\cite{Lou2007}. Spin valves with source-detector separations
between 250~nm and 2~$\mu$m were fabricated for each heterostructure
by electron beam lithography. In this case, the field is swept in
the plane, and the spin accumulation is inferred from the magnitude
of the jump $V_{\mathrm{NLSV}}$ in the voltage when the source and
detector magnetizations switch from parallel to antiparallel. Data
for $\mathrm{Co{}_{2}FeSi/\mathit{n}}$-GaAs at a source-detector
separation of 250~nm are shown in Fig.~\ref{fig:3}d. This spacing
is smaller than the spin diffusion length of approximately 800~nm
at room temperature (see Supplementary Fig.~4), and so it is possible
to measure $V_{\mathrm{NLSV}}$ over the entire temperature range
of this experiment. Figure~\ref{fig:3}e compares $V_{\mathrm{FMR}}$
with $V_{\mathrm{NLSV}}$ for both $\mathrm{Co{}_{2}FeSi/\mathit{n}}$-GaAs
and $\mathrm{Co{}_{2}MnSi/\mathit{n}}$-GaAs from 30~K to 300~K.
The temperature dependences are similar for both sets of measurements
on each heterostructure. Note that the choice of FM should not matter
in this case, as the temperature dependence is governed primarily
by properties of the SC. 

\begin{figure}[!h]
\includegraphics{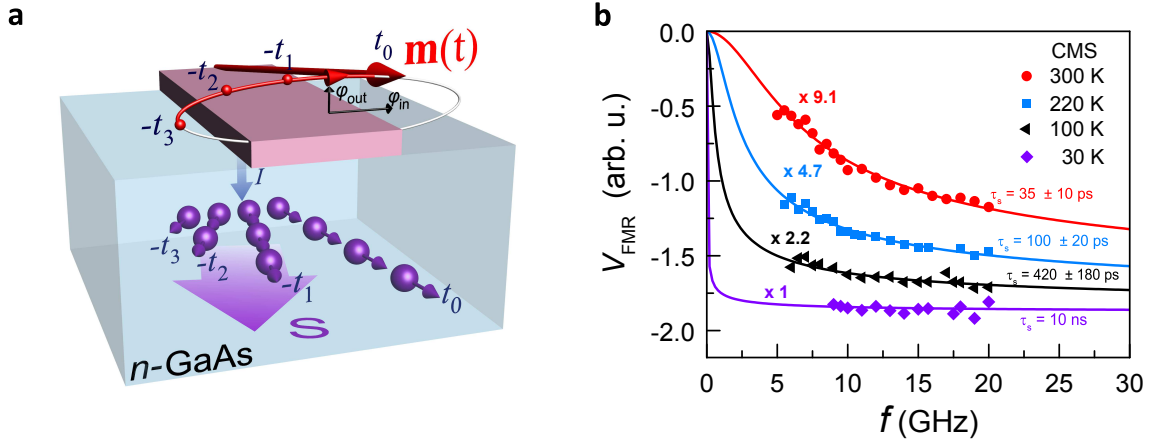}\protect\caption{\textbf{Illustration of the spin accumulation} \textbf{in the short
spin lifetime regime and the frequency dependence of $V_{\mathrm{FMR}}$
at different temperatures.} (\textbf{a}) Injected spins (small purple
arrows in $\mathit{n}$-GaAs) are parallel to the instantaneous orientation
of the magnetization (red arrow on top of the FM) at the time they
are injected. Earlier injected spins, e.g., at $t=-t_{1}$, $-t_{2}$
or $-t_{3}$ have progressively smaller magnitudes due to spin relaxation
and diffusion. The orientation of the average spin accumulation, indicated
by the large purple vector $\mathbf{S}$, lags behind the magnetization
$\mathbf{m}$, which decreases $\mathbf{\mathbf{\mathbf{s\cdot\hat{m}}}}$
relative to the static case ($\omega=0)$. (\textbf{b}) $\mathbf{\mathit{V}_{\mathrm{FMR}}}$
measured as a function of frequency at different temperatures. For
clarity, signal sizes are multiplied by the factors shown in the plot.
The solid lines are fits based on equation~(\ref{eq:4}) derived
in the main text, from which the electron spin lifetime in the SC
is obtained. For the 30 K data, the solid line is calculated from
the expression of $V_{\mathrm{FMR}}(\omega)$ by using the known spin
lifetime $\tau_{s}=10$~ns. \label{fig:4}}
\end{figure}

\noindent \textbf{Modeling and measuring the frequency dependence
to study spin relaxation.} The data of Fig.~\ref{fig:3} demonstrate
that $V_{\mathrm{FMR}}$ is proportional to the spin accumulation.
We now consider the frequency dependence of $V_{\mathrm{FMR}}$ and
its relationship to $\tau_{s}$.  Remarkably, we find that the FMR
technique can provide a measurement of the spin lifetime in the regime
$\omega_{\mathrm{FMR}}\tau_{s}\sim1$, which turns out to be accessible
at \emph{high }temperatures. In this limit, a spin injected from the
FM relaxes before the magnetization precesses through a complete cycle.
This physical scenario is illustrated in Fig.~\ref{fig:4}a, in which
the spin accumulation is represented by a series of spins (in purple)
that have been injected into the semiconductor at the times indicated,
where $t_{0}$ corresponds to the present, and $-t_{3}$ through $-t_{1}$
represent \emph{previous }times as labeled on the trajectory of the
precessing magnetization, which is shown in red. The spins shown are
oriented approximately parallel to the magnetization at the time at
which those spins were injected. Precession in the semiconductor is
ignored because $\omega_{L}<\omega_{\mathrm{FMR}}/10$ for the entire
frequency and magnetic field range covered by the experiment. The
smaller number of spins present from the earlier times is due to the
relaxation of spins over time. The average spin accumulation, shown
by the large purple arrow in Fig.~\ref{fig:4}a, is therefore a vector
which lags behind the precessing magnetization. 

\noindent 

The spin accumulation signal under these conditions can be calculated
by an appropriately weighted time average of equation~(\ref{eq:1}),
incorporating spin relaxation and diffusion\cite{Lou2006}:
\begin{equation}
V(t)=\eta P_{\mathrm{FM}}\dfrac{j_{s}}{e}\frac{\partial\mu}{\partial n}\mathbf{\mathrm{\intop_{-\infty}^{t}}}\frac{\hat{\mathbf{m}}(t)\mathbf{\cdot\hat{s}\mathrm{(\mathit{t'})}}}{\sqrt{2\pi D(t-t')}}\exp[-\frac{t-t'}{\tau_{s}}]dt',\label{eq:2}
\end{equation}
where $j_{s}$ is the injected spin current density, while $D$ and
$\tau_{s}$ are the spin diffusion constant and the spin lifetime
in \textit{$n$-}GaAs. Here $\hat{\mathbf{m}}(t)$ and $\mathbf{\hat{s}}\mathrm{(\mathit{t'})}$
represent the time-dependent orientations of the magnetization and
the injected spins, respectively. The integral over $t'$ evaluates
the projection of the spins injected at all previous times, reduced
by the effects of diffusion and relaxation, onto the present magnetization
direction $\hat{\mathbf{m}}(t)$. This is the \emph{instantaneous
}voltage detected by the FM. In the FM thin film, the precessing magnetization
traces out an ellipse, with an in-plane cone angle $\varphi_{\mathrm{in}}$
and an out-of-plane angle $\varphi_{\mathrm{out}}$. From geometric
analysis (ignoring the Larmor precession in the SC as noted in the
previous paragraph), $\hat{\mathbf{m}}(t)\mathbf{\cdot\hat{s}\mathrm{(\mathit{t'})}}$
takes the form

\noindent 
\begin{eqnarray}
 & \hat{\mathbf{m}}(t)\mathbf{\cdot\hat{s}\mathrm{(\mathit{t'})}}=1-\frac{\varphi_{\mathrm{in}}^{2}}{2}(\cos\omega t-\cos\omega t')^{2}\nonumber \\
 & -\frac{\varphi_{\mathrm{out}}^{2}}{2}(\sin\omega t-\sin\omega t')^{2},\label{eq:3}
\end{eqnarray}
(see Supplementary Note 3), in which $\omega$ is precession frequency
in the FM. The dc voltage $\left\langle V(t)\right\rangle $ in the
presence of a microwave field is the average value of $V(t)$ over
a precession period. The magnitude of $V_{\mathrm{FMR}}(\omega)$
is the difference of $\left\langle V(t)\right\rangle $ between resonance
$(\omega=\omega_{\mathrm{FMR}})$ and off resonance, which can be
determined by setting $\omega=0$ in equation~(\ref{eq:2}). We then
obtain
\begin{eqnarray}
 & V_{\mathrm{FMR}}(\omega)=\frac{1}{2}\eta P_{\mathrm{FM}}\frac{j_{s}}{eN(E_{\mathrm{f}})}(\varphi_{\mathrm{in}}^{2}+\varphi_{\mathrm{out}}^{2})\sqrt{\frac{\tau_{s}}{2D}}\nonumber \\
 & \times\left(\sqrt{\frac{1}{2\sqrt{1+\omega^{2}\tau_{s}^{2}}}+\frac{1}{2(1+\omega^{2}\tau_{s}^{2})}}-1\right).\label{eq:4}
\end{eqnarray}

\noindent From equation~(\ref{eq:4}) we obtain $\underset{\omega\rightarrow0}{\lim}V_{\mathrm{FMR}}(\omega)=0$.
This corresponds to the dc limit, in which the orientation of the
spin accumulation is always able to follow the magnetization. As $\omega$
increases, the angle by which the spin accumulation ``lags'' the
magnetization increases, and $V_{\mathrm{FMR}}(\omega)$ becomes more
negative. The expression for $V_{\mathrm{FMR}}(\omega)$ suggests
that in the regime $\omega\tau_{s}\sim1$, $V_{\mathrm{FMR}}(\omega)$
will have considerable frequency dependence. This regime corresponds
to short spin lifetimes, ie., $\tau_{s}\sim16$~ps for a typical
FMR frequency of $f=\omega/2\pi=10$ GHz. As $\omega$ increases further,
$V_{\mathrm{FMR}}(\omega)\rightarrow-\frac{1}{2}\eta P_{\mathrm{FM}}\frac{j_{s}}{eN(E_{\mathrm{f}})}(\varphi_{\mathrm{in}}^{2}+\varphi_{\mathrm{out}}^{2})\sqrt{\frac{\tau_{s}}{2D}}$
and saturates (no frequency dependence). This situation is equivalent
to the long $\tau_{s}$ limit which happens at low temperature. We
note that the magnitude of $V_{\mathrm{FMR}}(\omega)$ in this case
is proportional to the steady-state spin accumulation, with the proportionality
determined by the magnitude of the precession cone angles. This explains
the data presented in Figs~\ref{fig:2} and \ref{fig:3}.

To test this model, we measure the frequency dependence of $V_{\mathrm{FMR}}$
at different temperatures. During the measurement, the injection bias
current is fixed, and the amplitude of the magnetization precession,
which is measured from the TAMR at reverse bias\cite{Liu2014}, is
fixed at a constant value by adjusting the microwave power at each
frequency. Figure~\ref{fig:4}b presents the experimental results.
The FMR signal shows a pronounced frequency dependence at high temperatures.
The magnitude of the signal increases as the FMR frequency increases,
and the sign of the effect is negative as expected from the decrease
in $\mathbf{\left\langle \mathbf{\mathbf{s\cdot\hat{m}}}\right\rangle }$.
The solid lines in Fig.~\ref{fig:4}b are fits to $V_{\mathrm{FMR}}(\omega)$
given by equation~(\ref{eq:4}), except for the 30~K data, for which
$\tau_{s}$ is fixed at the value (10~nsec) inferred from non-local
spin valve measurements. Besides an overall scale factor, the spin
lifetime $\tau_{s}$ is the only fitting parameter ($D$ is absorbed
into the prefactor in equation~(\ref{eq:4})). We obtain a spin lifetime
of $\tau_{s}=35\pm10$~ps in $n$-GaAs at room temperature. This
value is comparable to previous theoretical calculations based on
D\textquoteright yakonov and Perel\textquoteright{} (DP) spin relaxation
in bulk GaAs\cite{Lau2001} as well as the value of $\tau_{S}$ ($50\pm10$~ps)
extracted from non-local spin valve measurements (see Supplementary
Fig.~4). The frequency dependence in Fig.~\ref{fig:4}b disappears
as the temperature is lowered to 30~K. In this limit, the magnetization
precesses through multiple cycles before the spin accumulation relaxes,
and $V_{\mathrm{FMR}}$ should therefore be insensitive to frequency,
as is found experimentally.

\noindent \textbf{\large{}Discussion}{\large \par}

\noindent In summary, we have demonstrated a new technique based on
FMR to measure spin accumulation in ferromagnet-semiconductor devices.
The precession of the magnetization at resonance leads to a measurable
phase lag between the spin accumulation in the SC and the magnetization
in the FM. Because the voltage detected by the FM is proportional
to the projection of the spin accumulation onto the magnetization,
the spin accumulation voltage decreases at resonance. The typical
FMR frequency is larger than the spin relaxation rate, and so the
spin accumulation can be measured up to room temperature, making this
approach more effective than the traditional Hanle effect, which has
previously limited spin lifetime measurements in GaAs-based spin transport
devices to temperatures less than approximately 150~K. FMR occurs
within a narrow magnetic field window, and the spin accumulation signal
detected by the FM is sensitive to the precession of the magnetization.
Together these make the technique essentially immune to the field-dependent
backgrounds that plague Hanle measurements, particularly at high temperatures.
By measuring the frequency dependence of the spin accumulation signal,
spin relaxation in the SC can be probed directly. With this method,
we obtain an electron spin lifetime of a few tens of picoseconds in
$n$-GaAs ($n=3\times10^{16}$$\mathrm{cm^{-3})}$ at room temperature.
The FMR-based spin detection technique developed here can be used
above room temperature and applied to other material systems, such
as metals, in which the spin lifetimes are short and traditional Hanle
measurements are impractical. 

\noindent \textbf{\large{}Methods}{\large \par}

\noindent \textbf{Sample growth.} The FM$/n$-GaAs heterostructures
investigated in this experiment were grown by molecular beam epitaxy
on GaAs (001) substrates. The growth started with a 500 nm undoped
GaAs buffer layer, followed by 2500~nm of Si-doped $n$-GaAs ($n$
= $3$$\times$$10^{16}$$\mathrm{cm^{-3}}$). The junction region
consists of a 15~nm $n\rightarrow n^{+}$-GaAs transition layer followed
by 18 nm $n^{+}$($5$$\times$$10^{18}$$\mathrm{cm^{-3}}$) GaAs\cite{Lou2007}.
The 5 nm thick FM film was then deposited epitaxially, followed by
10~nm thick Al and Au capping layers. The deposition temperatures
for Co$_{2}$MnSi and Co$_{2}$FeSi were $220^{\circ}\mathrm{C}$
and $270^{\circ}\mathrm{C}$, respectively. 

\noindent \textbf{Device fabrication.} Ion milling was used to define
the 5~$\mathrm{\mu m}$ $\times$ 50~$\mathrm{\mathrm{\mu}m}$ FM
contacts. The remote 50~$\mathrm{\mu m}$ $\times$ 50~$\mathrm{\mu m}$
non-magnetic contacts were fabricated by depositing 40~nm Cu and
then 40~nm Ge using electron beam evaporation. The CuGe contacts
were prepared in pairs and annealed by passing a current through them
in order to make them ohmic. This step avoids thermal annealing of
the FM$/n$-GaAs Schottky contact. A \textit{$n$}-GaAs channel was
defined by wet etching down to the substrate. 80~nm SiN was deposited
over the semiconductor using PECVD at a substrate temperature of $100^{\circ}\mathrm{C}$
and then lifted off in order to expose the CuGe and FM contacts. Finally,
Ti$/$Au electrodes and bonding pads were fabricated by depositing
25~nm Ti and 120~nm Au by electron beam evaporation.

\noindent \textbf{Measurement.} The FMR-spin detection measurement
was carried out using a microwave generator, a lock-in amplifier and
a current source, as shown in Fig.~\ref{fig:1}(a). The microwave
excitation signal was coupled to the FM injector through a coaxial
cable and an ordinary wire bond. A 120~nm thick Au wire patterned
over the FM contact functioned as a stripline, which was terminated
at ground. The dc current generating the spin accumulation was fed
to the sample using a conventional twisted pair, with one lead connected
to one of the ohmic contacts. The differential voltage between a second
ohmic reference electrode and the FM contact (connected through a
bias tee) was measured using a lock-in amplifier. The microwave signal
was modulated using the 100~Hz reference signal of the lock-in amplifier.
The circuit therefore measured the difference in the spin accumulation
with and without microwave excitation.

\noindent \textbf{Acknowledgments}

\noindent This work was supported by the National Science Foundation
(NSF) under DMR-1104951, C-SPIN, one of the six centers of STARnet,
a SRC program sponsored by MARCO and DARPA, the Materials Research
Science and Engineering Centers (MRSEC) program of NSF under DMR 08-19885,
and the NSF NNIN program.

\noindent \textbf{Author contributions }

\noindent C.L. fabricated the FMR devices, performed the FMR-spin
detection experiment, developed the theoretical model and wrote the
paper with P.A.C.. S.J.P. grew all the FM$/n$-GaAs heterostructures
used in this work. T.A.P. fabricated the non-local spin valve devices
and took the non-local spin valve measurement. C.C.G. and K.D.C. took
the non-local Hanle measurement. G.S. took some of the TAMR measurements.
C.C.G. developed software for performing the experiment and analyzing
the data. P.A.C. provided the initial concept, and P.A.C. and C.J.P.
supervised the project. 

\noindent \textbf{Additional information}

\noindent Competing financial interests: The authors declare no competing
financial interests.
\end{document}